\documentclass{emulateapj}

\slugcomment{draft of \today}
\shorttitle{Faraday Dispersion Functions of Galaxies}
\shortauthors{Ideguchi et al.}

\begin{document}
\title{Faraday Dispersion Functions of Galaxies}
\author{Shinsuke Ideguchi$^1$, Yuichi Tashiro$^1$, Takuya Akahori$^2$,
 Keitaro Takahashi$^1$, and Dongsu Ryu$^3$}
\affil{
$^1$University of Kumamoto, 2-39-1, Kurokami, Kumamoto 860-8555, Japan:  121d9001@st.kumamoto-u.ac.jp, 136d8008@st.kumamoto-u.ac.jp, keitaro@sci.kumamoto-u.ac.jp\\
$^2$Sydney Institute for Astronomy, School of Physics, The University of Sydney, NSW 2006, Australia: akahori@physics.usyd.edu.au\\
$^3$Department of Astronomy and Space Science, Chungnam National
University, Daejeon, Republic of Korea; ryu@canopus.cnu.ac.kr\\
}

\begin{abstract}

The Faraday dispersion function (FDF), which can be derived from an observed polarization spectrum by Faraday rotation measure synthesis, is a profile of polarized emissions as a function of Faraday depth. We study intrinsic FDFs along sight lines through face-on, Milky-Way-like galaxies by means of a sophisticated galactic model incorporating 3D MHD turbulence, and investigate how much the FDF contains information intrinsically. Since the FDF reflects distributions of thermal and cosmic-ray electrons as well as magnetic fields, it has been expected that the FDF could be a new probe to examine internal structures of galaxies. We, however, find that an intrinsic FDF along a sight line through a galaxy is very complicated, depending significantly on actual configurations of turbulence. We perform 800 realizations of turbulence, and find no universal shape of the FDF even if we fix the global parameters of the model. We calculate the probability distribution functions of the standard deviation, skewness, and kurtosis of FDFs and compare them for models with different global parameters. Our models predict that the presence of vertical magnetic fields and large scale-height of cosmic-ray electrons tend to make the standard deviation relatively large. Contrastingly, differences in skewness and kurtosis are relatively less significant.

\end{abstract}

\keywords{polarization --- galaxies: magnetic fields --- galaxies: structure --- techniques: polarimetric}

\section{Introduction}
\label{section1}

The origin and nature of magnetic fields in spiral galaxies is a longstanding problem \citep[see e.g.,][]{smk10}. Faraday rotation measure (RM) obtained by radio polarimetry is an important probe to examine structures of galactic magnetic fields \citep[e.g.,][]{gae05,bec09}. RM can be obtained as a proportionality constant between the polarization angle and the squared wavelength, if only a single polarization source is present along the line of sight (LOS). Inside galaxies, however, there are multiple sources in general. The relation thus becomes more complicated and is hard to be interpreted \citep{bu66,br05}. A more sophisticated method is needed to be developed.

Faraday RM synthesis \citep{bu66,br05} is an alternative method to interpret radio polarimetry data. Using Faraday RM synthesis, we can transform an observed polarization spectrum into the so-called Faraday dispersion function (FDF), which includes information of polarized sources along the LOS. Faraday RM synthesis has been applied only recently, since a wideband polarization spectrum is needed to ensure reasonable transformation into the FDF. Up to now, there are several observations of FDFs in Galactic interstellar medium \citep{sch09}, galaxies \citep{hea09,gov10,wol10}, galaxy clusters \citep{db05,br11,piz11}, and active galactic nuclei \citep{osu12}. For example, \cite{hea09} studied FDFs of the WSRT-SINGS galaxies and found multiple nuclear components that are offset to both positive and negative RM by $100-200$~${\rm rad/m^{-2}}$. \cite{db05} and \cite{br11} studied the Perseus cluster field with the WSRT and tried to separate components of the cluster and the Milky Way in RM between $-50$ and $+100$~${\rm rad/m^{-2}}$. In future, the Square Kilometre Array (SKA) and its pathfinders will enable ultra-wideband observations and the FDF will become a key observable quantity for the study of cosmic magnetism in the next decades.

Galactic polarized radio emissions are mostly contributed from synchrotron radiation by cosmic-ray electrons, so that the FDF reflects distributions of cosmic-ray electrons as well as thermal electrons and magnetic fields which determine RM. Thus, FDFs of galaxies are expected to be a new probe that contains rich information about internal structures of galaxies. An uncertainty of the probe is complexity of the FDF. As reported in the Milky Way, a small volume filling factor of electrons suggests their clumpy distribution \citep{ber06, hil08, gmcm08}. Highly disturbed distributions of polarization angle and RM indicate the existence of turbulent magnetic fields \citep{sun08, wae09, gae11, bur12}. These would result in complicated shape of FDF. So far, there are a number of theoretical works which investigated the capability of Faraday RM synthesis \citep[see][references therein]{sun14}. But most of them have considered a simple form of FDF such as the delta function, top hat, and gaussian function for simplicity.

There are some attempts which considered turbulence on the FDF of galaxies using simple models \citep[e.g.,][]{bel11,fri11,bec12}. For instance, \cite{bel11} took into account a simple magnetic field reversals, more or less realistic galactic model, and simple Gaussian random magnetic fields, and then discussed Faraday caustics -- sharply peaked and asymmetric profiles are shown in the FDF because of LOS magnetic-field reversals. They examined observed FDFs assuming observations by radio telescopes such as the LOw-Frequency ARray (LOFAR) and Australian SKA Pathfinder (ASKAP) POSSUM survey.

Although previous studies have indicated an importance of simulating observed FDFs, it is complementary to focus on intrinsic FDFs in detail. Since real FDFs of galaxies would be complicated, it is meaningless to study observed FDFs without confirming by simulations that intrinsic FDFs of galaxies can possess meaningful information of galaxies and the information can be interpreted from at least intrinsic FDFs. If the information is not available from intrinsic FDFs, it is impossible to obtain the information from observed FDFs. A simple model such as global magnetic-fields only and/or Kolmogorov-like turbulent magnetic-fields does not allow us to address this question, and a study of intrinsic FDFs based on realistic galactic models is necessary.

In this paper, we investigate intrinsic FDFs of galaxies in detail with a latest galactic model \citep{arkg13} which based on observational data of the Milky Way and incorporates three dimensional magneto-hydrodynamic (MHD) turbulence. Observed FDFs are not simulated in this paper. We demonstrate how properties of a galaxy are reflected into the intrinsic FDF, and show some statistical quantities to characterize global shapes of FDFs for different galactic models. In section \ref{section:2}, we describe the basic idea of Faraday RM synthesis and method to calculate the FDFs using galactic models. The resultant FDFs and their characterization are presented in section \ref{section:3}. Discussion and summary of our results follow in section \ref{section:4}.

\section{Calculation Method}
\label{section:2}

\subsection{Faraday RM Synthesis}
\label{subsection:2.1}

In this section, we explain Faraday RM synthesis first proposed by \cite{bu66} and extended by \cite{br05}. The complex polarized intensity $P(\lambda^2)$ as a function of wavelength $\lambda$ is expressed as,
\begin{eqnarray}\label{eq:PI}
P(\lambda^2)
&=& p(\lambda^2)I(\lambda^2) \nonumber \\
&=& Q(\lambda^2)+iU(\lambda^2) \nonumber \\
&=& \int_{-\infty}^{+\infty}F(\phi)e^{2i\phi\lambda^2} d\phi,
\end{eqnarray}
where $p$ is the complex fractional polarization, and $I$, $Q$, and $U$ are the Stokes parameters. $F(\phi)$ is the Faraday dispersion function (FDF), which is a complex polarized intensity as a function of Faraday depth $\phi$. The Faraday depth (or the RM when the FDF consists of a single delta function) is defined as 
\begin{equation}\label{eq:FD}
\phi(x)
= 810 \int_x^0 \left( \frac{n_{\rm e} (x')}{1~{\rm cm}^{-3}} \right)
               \left( \frac{B_\parallel(x')}{1~\mu {\rm G}} \right)
               \left( \frac{dx'}{1~{\rm kpc}} \right) ~ {\rm rad/m^2},
\end{equation}
where $n_{\rm e}$ is the electron density, $B_\parallel$ the LOS component of magnetic fields, $x'$ the physical distance, and $x$ is the distance to a source from an observer at $x'=0$. 

The FDF can be reconstructed by the inverted equation of Eq. (\ref{eq:PI}):
\begin{equation}\label{eq:FDF}
F(\phi)
=\int_{-\infty}^{+\infty}P(\lambda^2)e^{-2i\phi\lambda^2}d\lambda^2.
\end{equation}
This transform is called Faraday RM synthesis, which gives the FDF from observed $P(\lambda^2)$.

It should be noted that there is, in general, no one-to-one correspondence between $\phi$ and $x$. The FDF thus does not straightforwardly give the distribution of polarized intensity in physical space. Nevertheless, the distribution of magnetic fields and thermal and cosmic-ray electrons are reflected in the FDF and it is possible to extract information on distribution \citep{bec12}.

\subsection{Model of Galaxies}
\label{subsection:2.2}

In order to study FDFs of galaxies, we employ a model of the Milky Way galaxy constructed by \cite{arkg13}. Actually, this is a model of the Solar neighbor but we use it as a representative model because physical quantities have been better determined there. We expect that the model are reasonable also for external spiral galaxies. Below, we briefly summarize the model.

The model considers the domain toward high Galactic latitudes of the Milky Way. It consists of the global, regular component as well as the turbulent, random component. The regular component is modeled using the thermal electron density model of \cite{cl02} and magnetic-field models including an axi-symmetric spiral field and a halo toroidal field \citep{sun08} plus a dipole poloidal field that produces a coherent vertical field near the Earth \citep{gkss10}. The random component is modeled by piling up computational boxes of MHD turbulence simulations \citep{krjh99} from the Galactic mid-plane up to $\pm 10.0$ kpc, where each box size is $500~{\rm pc}$ with $512$ grids on a side. They adopted a driving scale of $250~{\rm pc}$ and a rms flow speed of $30~{\rm km/s}$ based on H$\alpha$ observations \citep{hil08}. Then, Faraday depth is calculated with the above models.

Over the frequencies we are considering ($\sim 100~{\rm MHz}-10~{\rm GHz}$), polarized emissions of galaxies mostly comes from synchrotron radiation of cosmic-ray electrons, and other radiation mechanisms can be safely neglected. This means that we need information of a cosmic-ray electron density in addition to magnetic fields derived from the above model. For it, we utilize an exponential model based on observations \citep{sun08}. With the cosmic-ray electron density and the amplitude and direction of magnetic fields, synchrotron emissivities and specific Stokes parameters are calculated following Equations (3-5) of \cite{wae09}. In this paper, we calculate the Stokes parameters at 1 GHz.

\subsection{Calculation}
\label{subsection:2.3}

We suppose to observe a face-on spiral galaxy and use Cartesian coordinates $(x,y,z)$, where the $x-y$ plane coincides with the galactic mid-plane with $y$ pointing the anti-galactic center and $z$ penetrates the mid plane. The LOS is set to be parallel to $z$ axis, and the field-of-view center directs to $(x,y)=(0,8.5)$ in kpc (the location of the Earth in the Milky Way). We only consider the computational region of $-0.25 < x < 0.25$, $8.25 < y < 8.75$, and $-10 < z < 10$ all in kpc. The simulated $500~{\rm pc}\times 500~{\rm pc}$ region corresponds to $\sim 10''\times 10''$ if we put the galaxy at a distance of 10 Mpc from the Earth (e.g., at Virgo cluster).

We divide the computational region by 32 grids each in $x$ and $y$, and by 1280 grids in $z$. For each grid, we calculate the Faraday depth using Equation (\ref{eq:FD}), and calculate the polarized intensity described in Section~\ref{subsection:2.2}. We integrate the intensity along the LOS and then we obtain the FDF by sorting the intensity as a function of Faraday depth. We also obtain Faraday depth as a function of $z$. The Faraday depth for a grid cell is calculated at the middle of the cell, and the polarized intensity of a cell is an average within the cell. We set the resolution in Faraday depth for the calculation of FDFs to $0.1~{\rm rad/m^2}$, which resolution is sufficiently high to follow the variation of the FDF, so that the results do not change significantly with higher resolution in Faraday depth.

In \cite{arkg13}, the model parameters which determine global configurations of the galaxy such as the scale heights of thermal and cosmic-ray electron densities and mean strength of coherent vertical magnetic fields were chosen based on observations of the Milky Way. In this work, we can vary the values of such global parameters because our purpose is not to reproduce observed galaxies but to demonstrate how galaxies look like in terms of FDF. The above set of parameters are curious for studying the intrinsic FDFs since they are all fundamental ones of galaxies, and major ones that mainly characterize the FDF for face-on galaxies. Even some extreme cases would be helpful to understand the nature of the FDF. We take the "ADO model" in \cite{arkg13} as our fiducial model and consider three variants summarized in Table \ref{table1}. Considerations for some other parameters will be commented in Section \ref{section:4}.

%%% table1 here %%%
\begin{deluxetable}{cccc}
\tablenum{1}
\tabletypesize{\footnotesize}
\tablewidth{0pt}
\tablecaption{Models of galaxies.\label{table1}}
\tablehead{
Model & vertical & cosmic-ray electron  & thermal electron \\
& magnetic field& scale height & scale height \\
& $B_{{\rm p},z}$ ($\mu$G) & $h_{\rm c}$ (kpc) & $h_1$ (kpc)
}
\startdata
1 & 0 & 1.0 & 1.0  \\
2 & $\sim 1.0$ & 1.0 & 1.0 \\
3 & 0 & 3.0 & 1.0 \\
4 & 0 & 1.0 & 3.0
\enddata
\end{deluxetable}

The first parameter we vary is the mean strength of vertical magnetic field. Here, the random vertical magnetic field always exists in the model, and this component additionally incorporates the global, regular magnetic field along $z$ axis. The component is modeled with the galactic-center dipole field, where the field is penetrating the galactic plane, and is written as \citep{gkss10}
\begin{equation}\label{eq:Bp}
B_{{\rm p},z}(z) =\mu_{\rm p}(1-3(z/r)^2)/r^3,
\end{equation}
where $r=(x^2+y^2+z^2)^{1/2}$. For Model 2, we set $\mu_{\rm p}$ to a value such that the strength $B_{{\rm p},z}(z)\sim 1.0~{\rm \mu G}$ around $(x,y)=(0,8.5)$ in kpc while it is absent for other models. The second parameter is the scale height of cosmic-ray electron density. The cosmic-ray electron density is written as \citep{sun08}
\begin{equation}\label{eq:Cp}
C(z) \propto C_0 \exp(-|z|/h_{\rm c}),
\end{equation}
where $C_0=0.0001$~${\rm cm^{-3}}$ and the scale height $h_{\rm c}=1$~kpc for Models 1, 2 and 4 and $3$~kpc for Model 3. The third parameter is the scale height of thermal electron density of the thick disk component, which is given by \citep{cl02}
\begin{equation}\label{eq:ne}
n_1(z) \propto n_1 {\rm sech^2}(z/h_1),
\end{equation}
where $n_1=0.02$ ${\rm cm^{-3}}$ and the scale height $h_1=1$~kpc for Models 1, 2 and 3 and $3$~kpc for Model 4.

\section{Result}
\label{section:3}

\subsection{Complexity of FDF}
\label{subsection:3.1}

\begin{figure}[tp]
\epsscale{1.0}
\plotone{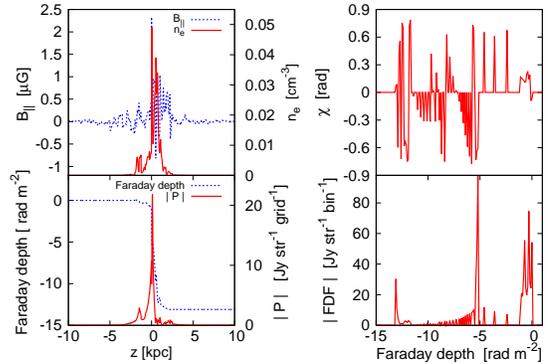}
\caption{
Top-left: LOS ($z$-axis) distributions of thermal electron density ($n_e$, the red solid line) and LOS component of magnetic fields ($B_{||}$, the blue dotted line). Bottom-left: LOS distributions of the absolute value of polarized intensity ($|P|$, the red solid line) and Faraday depth (the blue dashed line). Top-right: Faraday depth ($\phi$) distribution of the average polarization angle per bin of $\Delta \phi = 0.1$. Bottom-right: Faraday depth distribution of absolute value of Faraday dispersion function (FDF) per bin of $\Delta \phi = 0.1$.}
\label{fig:f1}
\end{figure}

\placefigure{f1}

In this subsection, we show some examples of FDFs of galaxies of Model 1. First, we calculate FDFs of a single grid in $x$-$y$ plane ($16~{\rm pc} \times 16~{\rm pc}$) which corresponds to a beam size of $0.3''\times 0.3''$ at 10 Mpc.

Figure \ref{fig:f1} shows the result for a relatively simple case where the Faraday depth is virtually a monotonic function of physical distance. The top-left panel shows the distributions of thermal electron density and LOS ($z$ axis) component of magnetic fields in physical space along a LOS. As explained in the previous section, the density is concentrated near the galactic mid-plane within the scale height ($|z| \lesssim 1~{\rm kpc}$), while magnetic fields extend more to outer regions due to the halo magnetic fields. Both density and magnetic-field distributions have violent fluctuations caused by turbulence.

The resultant Faraday depth is shown as the blue dotted line in the bottom-left panel of Figure \ref{fig:f1}. In this case, Faraday depth almost monotonically decreases with the physical distance because magnetic fields are mostly positive. The panel also shows the polarized intensity distribution, which shows that the intensity is peaked at the mid plane with tails in the halo. We find that variations of Faraday depth in halo ($|z| > 1~{\rm kpc}$) is small ($<1~{\rm rad/m^{2}}$), even though sub $\mu$G magnetic field extends to several kpc scale. This indicates the feature of Faraday caustics that polarized emissions in the halo tend to contribute to the FDF of the same value of a Faraday depth. 

The right panels show the polarization angle (top) and the FDF (bottom). We can see that the absolute value of the FDF has a main peak at $\phi = -5 ~{\rm rad/m^2}$ and sub peaks at $\phi = -13$ and $-1~{\rm rad/m^2}$ with different polarization angles each other. The main and two sub peaks can be attributed to the emissions in the mid plane at $z\sim 0$, the halo with $z>0$, and the halo with $z<0$, respectively, as implied by the relation between $\phi$ and $z$ (the bottom-left panel). As expected, although polarized intensities are small in the halo, we can see them in the FDF as large peaks because they are accumulated to relatively large values at $\phi = -13$ and $-1~{\rm rad/m^2}$.

\begin{figure}[tp]
\epsscale{1.0}
\plotone{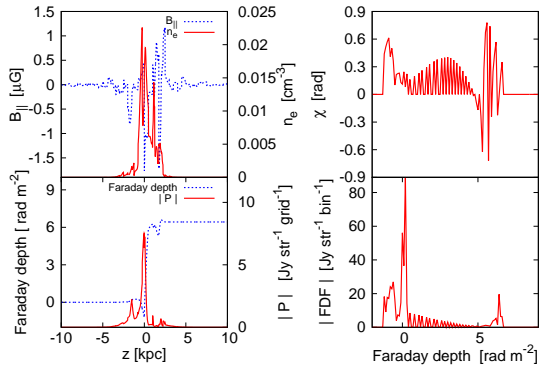}
\caption{Same as Figure \ref{fig:f1} but for a non-monotonic, relatively-simple Faraday depth.}
\label{fig:f2}
\end{figure}

\placefigure{f2}

Since the sign of the LOS component of magnetic fields can change, in general, Faraday depth does not vary monotonically with the physical distance. Figure \ref{fig:f2} represents a less simple case. Again, the bottom-left panel shows that the polarized intensity is peaked at the mid-plane with tails in the halo, but the behavior of the Faraday depth is no longer monotonic with respect to physical distance. 

As is shown in the bottom-right panel of Figure \ref{fig:f2}, there are three peaks at $\phi = -1, 0$ and $7~{\rm rad/m^2}$ in the FDF, similar to the previous case in Figure~\ref{fig:f1}. However, the main peak at $\phi = 0~{\rm rad/m^2}$ is attributed to the emissions in the halo with $z<0$ (see the bottom-left panel), rather than the mid plane at $z\sim 0$. The emissions in the mid plane and in the halo with $z>0$ appear in the FDF at $\phi = -1$ and $7~{\rm rad/m^2}$, respectively. Therefore, the order of peak locations is inverted in the FDF compared with that in physical space, because of the non-monotonicity of the Faraday depth. Moreover, peak amplitudes of the FDF does not straightforwardly suggest the origin of the emissions (mid-plane or halo), due to their degeneracies in a certain Faraday depth.

\begin{figure}[tp]
\epsscale{1.0}
\plotone{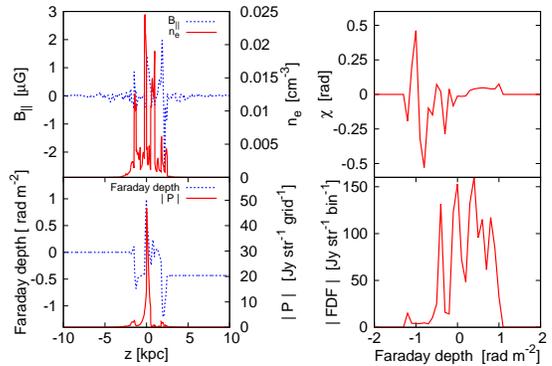}
\caption{Same as Figure \ref{fig:f1} but for a general complicated Faraday depth.}
\label{fig:f3}
\end{figure}

\placefigure{f3}

Figure \ref{fig:f3} shows a case with an Faraday depth which is far from monotonic. It is seen that the FDF is limited to a region with small values of $|\phi|$ compared with the previous two cases. This is because the Faraday depth behaves like random-walk and does not reach large values. The relation between $\phi$ and $z$ is thus far from the one-to-one correspondence. The behavior of the Faraday depth is so complicated that it is difficult to find the correspondence between the peaks in the FDF and those in physical space. This is the generic situation and simple behaviors of the two cases in Figures \ref{fig:f1} and \ref{fig:f2} are rather exceptional. Thus, as we saw above, the FDF look very different even for a fixed set of global parameters and is highly dependent on the different realizations of turbulent components which determine the behavior of Faraday depth.

\begin{figure}[tp]
\epsscale{1.0}
\begin{center}
\plotone{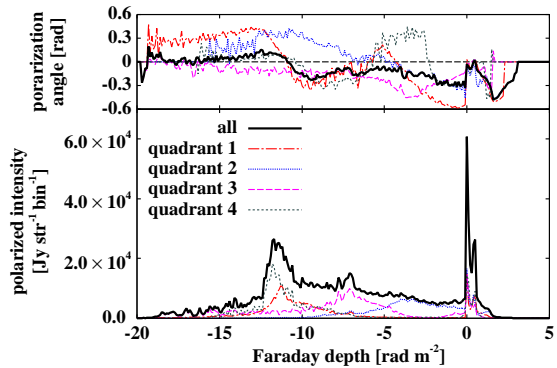}
\end{center}
\caption{The average polarization angle (upper) and the absolute value of Faraday dispersion function (lower) integrated over the whole simulated region ($500~{\rm pc} \times 500~{\rm pc}$). Contributions from four quadrants ($250~{\rm pc} \times 250~{\rm pc}$) are also plotted.}
\label{fig:f4}
\end{figure}

\placefigure{f4}

We also calculate FDFs of the whole region in x-y plane ($500~{\rm pc} \times 500~{\rm pc}$) which corresponds to a beam size of $10'' \times 10''$. Figure \ref{fig:f4} shows the FDF integrated over this region as well as those of the four quadrant ($250~{\rm pc} \times 250~{\rm pc}$). We see that the integrated FDF is much broader and smoother compared with those in Figures \ref{fig:f1} - \ref{fig:f3}. Such broadening and smoothing is ascribed to the fact that each quadrant contributes to a different part of the total FDF, shown as color lines. The result demonstrates that an observation with a larger beam size produce a broader and smoother FDF, in which various FDFs having peaks at different Faraday depths are mixed together.

Note that the absolute value of the total FDF is not in general equal to the sum of the absolute values of FDFs for the quadrants because the polarization angles are different each other. This is the effect so-called beam depolarization, though it is not so significant at the shown scales because coherent magnetic fields dominate over the turbulent magnetic fields.

\subsection{FDF and the global properties of galaxy}
\label{subsection:3.2}

If the FDF of galaxies can be described with simple functions, we can quantify the FDF easily. However, as shown in the previous subsection, FDFs of galaxies are very complicated in general and it is very difficult to translate them into the distribution of physical quantities in physical space. Nevertheless, it would be possible to extract some global properties of galaxies, such as the average amplitude and the coherence length of magnetic fields and the scale heights of thermal and cosmic-ray electrons. This is because such properties could affect the FDF through the behavior of the Faraday depth as a function of physical distance. Below, we consider statistical properties of FDFs of the whole region for 4 models in Table \ref{table1}, mainly for the purpose of interpreting the global shape of the FDFs.

\begin{figure}[tp]
\epsscale{0.9}
\begin{center}
\plotone{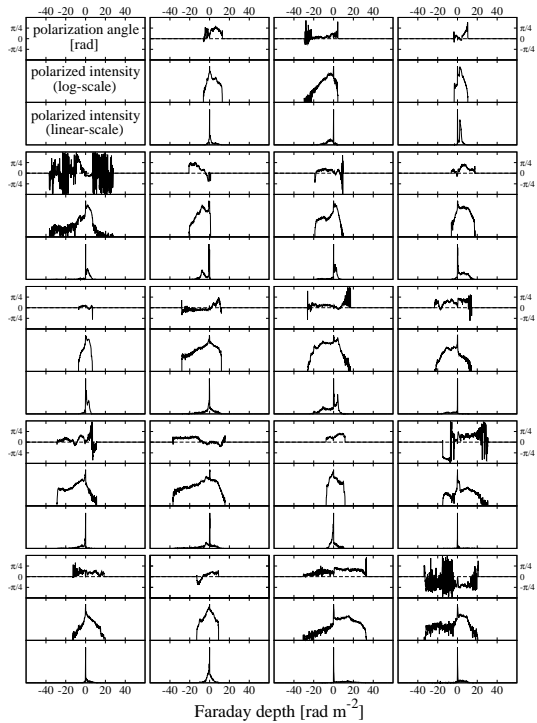}
\end{center}
\caption{
Results of 19 out of 800 runs for Model 1. Each has the same global parameters given in Table \ref{table1} but different configuration of turbulence. For each run, top panel shows the polarization angle, middle panel shows the Faraday dispersion function (FDF) in the linear scale, and the bottom panel shows the FDF in the log scale. FDFs are shown in arbitrary unit for clear display.
}
\label{fig:f5}
\end{figure}

\placefigure{f5}

\begin{figure}[tp]
\epsscale{0.9}
\plotone{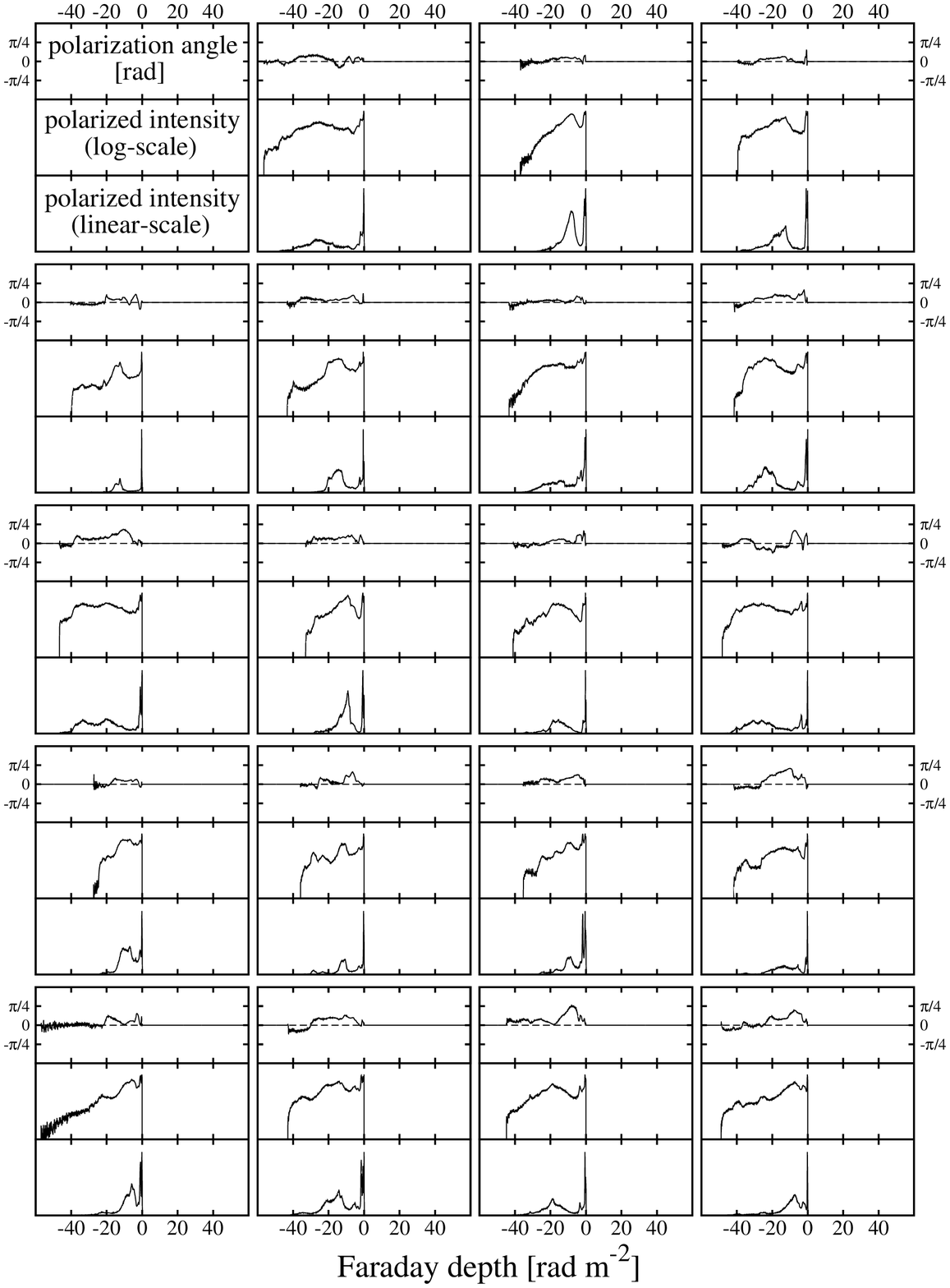}
\caption{Same as Fig.\ref{fig:f5} but for Model 2.}
\label{fig:f6}
\end{figure}

\placefigure{f6}

\begin{figure}[tp]
\epsscale{0.9}
\plotone{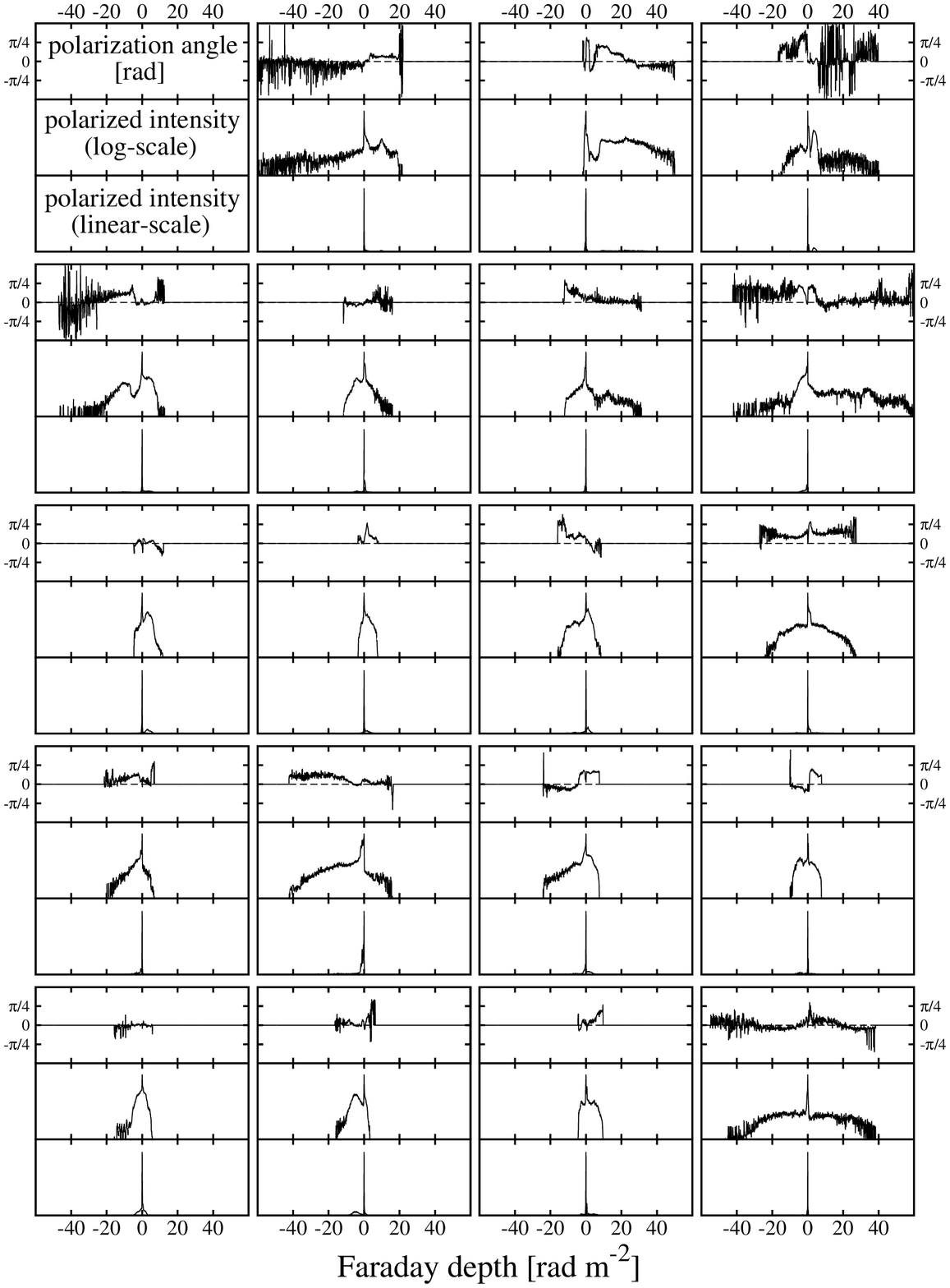}
\caption{Same as Fig.\ref{fig:f5} but for Model 3.}
\label{fig:f7}
\end{figure}

\placefigure{f7}

\begin{figure}[tp]
\epsscale{0.9}
\plotone{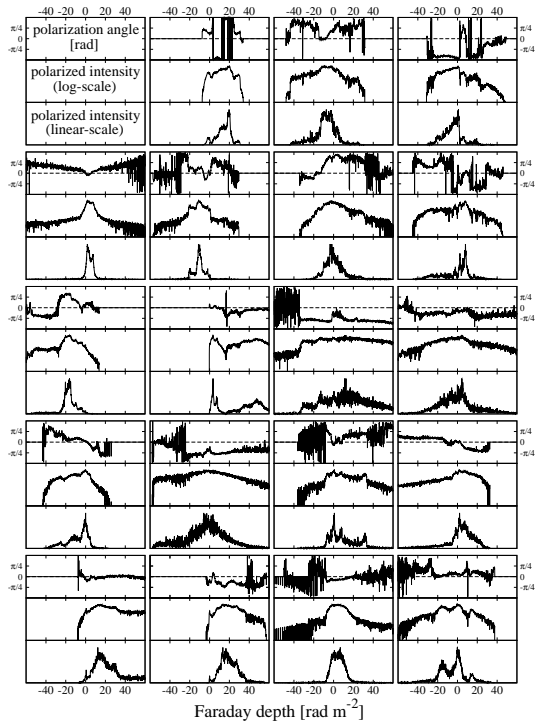}
\caption{Same as Fig.\ref{fig:f5} but for Model 4.}
\label{fig:f8}
\end{figure}

\placefigure{f8}
We carried out 800 runs with different realizations of turbulence for each models. To do this, we randomly shift and rotate computational boxes of MHD turbulence simulations when we pile up them. Figure \ref{fig:f5} shows 19 examples of FDFs for Model 1. We find that the shape of FDF varies significantly for different configurations of turbulence. For example, some FDFs have a single thin peak and others have multiple peaks. Such variations can be also seen for Models 2 to 4 (Figures \ref{fig:f6} to \ref{fig:f8}). Therefore, we see no universal shape of the FDF, even if we fix the parameters characterizing the global properties mentioned above.

Apparently, Models 2 (Figure~\ref{fig:f6}) and 4 (Figure~\ref{fig:f8}) have broader FDFs and this feature may allow us to probe some global properties of galaxies. Since the reconstructed FDF tends to show a peaked profile with tails, to quantify the features of the global shape of FDFs, we calculate the standard deviation $\sigma$, skewness $\gamma_{\rm s}$ and kurtosis $\gamma_{\rm k}$ of FDFs, defined as,
\begin{eqnarray}
\sigma^2 &=& \frac{\sum_i |F(\phi_i)| (\phi_i-\mu)^2}{\sum_i |F(\phi_i)|} \\
\gamma_{\rm s} &=& \frac{\sum_i |F(\phi_i)| (\phi_i-\mu)^3}{\sigma^3 \sum_i |F(\phi_i)|} \\
\gamma_{\rm k} &=& \frac{\sum_i |F(\phi_i)| (\phi_i-\mu)^4}{\sigma^4 \sum_i |F(\phi_i)|},
\end{eqnarray}
where $\phi_i$ is the Faraday depth of the $i$-th bin and,
\begin{equation}
\mu = \frac{\sum_i |F(\phi_i)| \phi_i}{\sum_i |F(\phi_i)|},
\end{equation}
is the average Faraday depth. We note that we take the definition for the kurtosis so as to the gaussian function equals to 3. We calculate the probability distribution functions of these quantities, using 800 FDFs with different configuration of turbulence, for each model.

\begin{figure}[tp]
\epsscale{1.0}
\plotone{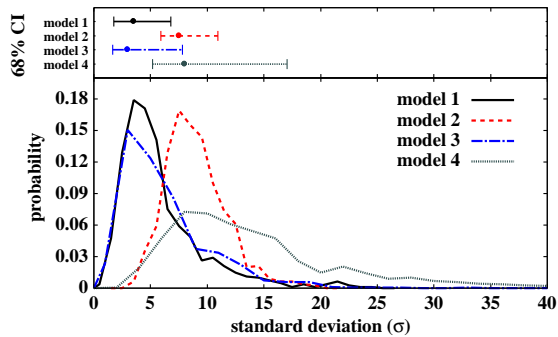}
\caption{Probability distribution functions (lower panel) and 68\% confidence intervals (CI, upper panel) for the standard deviation of Faraday dispersion functions for the four models.}
\label{fig:f9}
\end{figure}

\placefigure{f9}

\begin{figure}[tp]
\epsscale{1.0}
\plotone{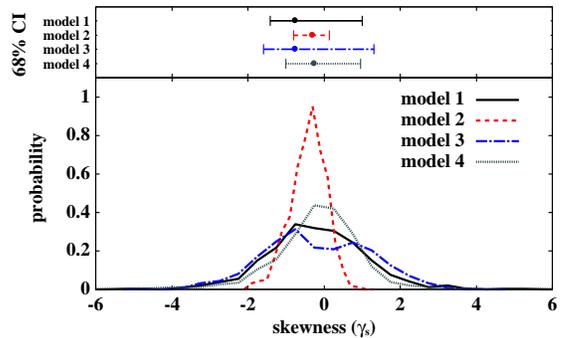}
\caption{Same as Fig. \ref{fig:f9} but for skewness.}
\label{fig:f10}
\end{figure}

\placefigure{f10}

\begin{figure}[tp]
\epsscale{1.0}
\plotone{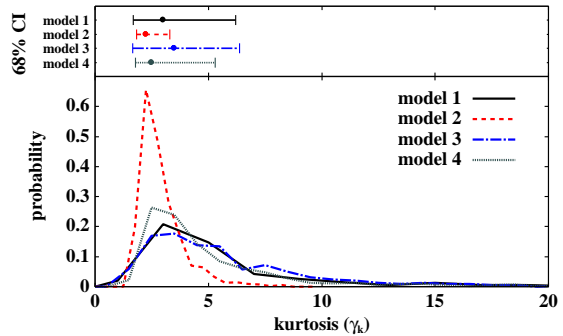}
\caption{Same as Fig. \ref{fig:f9} but for kurtosis.}
\label{fig:f11}
\end{figure}

\placefigure{f11}

The probability distribution functions and the 68\% confidence intervals for the standard deviation, skewness, and kurtosis are shown in Figures \ref{fig:f9} - \ref{fig:f11}, respectively. The most obvious difference can be seen in the probability distribution of the standard deviation, where Models 2 and 4 have larger values. This can be ascribed to the fact that the Faraday depth tends to reach large absolute values, although the physical reason is different between Model 2 and Model 4. As to Model 2, it is due to the presence of the coherent vertical magnetic field which pushes up the Faraday depth to large values. Contrastingly, as to Model 4, the Faraday depth behaves like random walk but the extended thermal electrons increase the expectation value of the variance of the Faraday depth.

On the other hand, although 68\% confidence intervals are mostly overlapped with each other in the probability distribution of skewness and kurtosis, for Model 2, these quantities are limited to relatively narrow ranges. The skewness for Model 2 tends to be negative and this is due to the orientation of coherent vertical magnetic fields.

\section{Summary and Discussion}
\label{section:4}

In this paper, we calculated the Faraday dispersion function (FDF) of face-on spiral galaxies, which represents the distribution of the polarized intensity as a function of Faraday depth, using realistic galactic models \citep{arkg13} which are based on observational data and numerical simulations of turbulence. FDFs reflect the distribution of magnetic fields and thermal and cosmic-ray electrons and can be a useful tool to probe galaxies. However, in general, FDFs do not have one-to-one correspondence with the distribution of the polarized intensity in physical space and are not easy to give a direct interpretation. Further, the shape of FDFs varies significantly for different configurations of turbulence, even if we fix the global parameters of the galactic model such as the mean amplitude and correlation length of magnetic fields and the scale heights of thermal and cosmic-ray electron number densities. We, especially, focused on the intrinsic FDFs rather than the observed FDFs to investigate how much information about the parameters the intrinsic FDFs have. It is necessary to interpret intrinsic FDFs, since it is impossible to obtain some information from observed FDFs if the information is not available from intrinsic FDFs.

Then, as simple measures to characterize the global shape, we calculated the probability distribution functions of the standard deviation, skewness and kurtosis of the FDFs and compared them for different models. We found that the presence of vertical magnetic fields and large scale-height of cosmic-ray electrons make the standard deviation relatively large so it can be a useful indicator of them. However, it should be noted that the effect of vertical fields would become smaller when we consider spiral galaxies with inclination. Contrastingly, differences in skewness and kurtosis are not so significant. Obviously, we need more sophisticated measures for the characterization of FDFs and we will pursue it in near future.

In the current study, the calculation of the FDF is limited to a relatively small region with 500 pc $\times$ 500 pc. At this scale, magnetic fields perpendicular to the LOS are dominated by global coherent fields rather than turbulent fields. This is why the beam depolarization is not so significant. It would be interesting to extend our calculation to larger regions to study the possibility of probing the global topology of magnetic fields by combining the Faraday tomography and imaging. Remembering that FDFs are contributed appreciably from the halo region, as well as the disk, halo fields would also be able to be probed. For instance, the three peaks seen in the FDF would not be expected if there are no halo toroidal magnetic field, although the three peaks do not always appear in the FDF even if there are halo toroidal magnetic field.

We also compared the cases between axi-symmetric and bi-symmetric spiral fields, and the cases between the dipole poloidal field and the X-field \citep{arkg13}, but we did not see notable differences on the resultant FDFs. This may be due to the fact that we calculated FDFs for a relatively small region up to 500 pc $\times$ 500 pc. A study for probing the global topology of magnetic fields could be possible by using global MHD simulations of galactic gaseous disk \citep{mac13} or by improving the model we adopted. For the improvement, further investigations of turbulence such as a Mach number, the plasma beta, the driving scale, and so on, at mid Galactic latitudes are necessary.

Let us discuss the observational prospect for galactic FDFs. In the past studies, the shape of galactic FDFs has often been assumed to be delta function, top-hat or gaussian function. However, we saw that realistic FDFs are not so simple and even a single galaxy has multiple peaks. Furthermore, it is common to assume the intrinsic polarization angle does not change within a single source, which is not obviously reasonable. These issues are particularly important when we use QU-fitting method \citep{osu12,ide14}, where we need to assume a specific functional form of FDF and fit it with the data, $P(\lambda^2)$, to obtain the parameter values. We need further studies to find appropriate functions to be fitted.

On the other hand, with RM synthesis, we can directly reconstruct the FDF from the data without any assumption on its form. However, the reconstruction is not perfect because $P(\lambda^2)$ is not physical for negative $\lambda^2$ and is observationally limited even for positive $\lambda^2$. Especially, it was shown that multiple peaks tend to induce false signal in FDF \citep{far11,kum14}. It is important to study how well the realistic FDFs can be reconstructed and how precisely the characteristic quantities such as the dispersion, skewness and kurtosis can be obtained by RM synthesis and other sophisticated methods \citep{and11,fri11,bec12,aka13}. We will pursue these issues in near future.

\medskip

\acknowledgments
K.T. is supported in part by the Grant-in-Aid from the Ministry of Education, Culture, Sports, Science and Technology (MEXT) of Japan, No. 23740179, No. 24111710 and No. 24340048. T.A. acknowledges the supports of the Japan Society for the Promotion of Science (JSPS). D.R. acknowledges the supports of National Research Foundation of Korea through grant 2007-0093860.

%%% \clearpage


\begin{thebibliography}{}
\bibitem[Akahori et al.(2013a)]{aka13}
Akahori, T., Kumazaki, K., Takahashi, K. \& Ryu, D. 2013, submitted
\bibitem[Akahori et al.(2013b)]{arkg13}
Akahori, T., Ryu, D., Kim, J., \& Gaensler, B. M. 2013, \apj, 767, 150
\bibitem[Andrecut et al.(2011)]{and11}
Andrecut, M., Stil, J. M., \& Taylor, A. R. 2011, \apj, 143, 33
\bibitem[Andrecut (2013)]{and13}
Andrecut, M. 2013, \mnras, 430, L15
\bibitem[Beck(2009)]{bec09}
Beck, R. 2009, Rev. Mexicana Astron. Astrofis. (Serie de Conferencias), 36, 1
\bibitem[Beck et al.(2012)]{bec12}
Beck, R., Frick, P., Stepanov, R. \& Sokoloff, D. 2012, \aap, 543, A113
\bibitem[Bell et al.(2011)]{bel11}
Bell, M. R., Junklewitz, H., \& Ensslin, T. A. 2011, \aap 535, A85
\bibitem[Bell et al.(2013)]{bel13}
Bell, M. R., Oppermann, N., Crai, A., \& Ensslin, T. A. 2013, \aap 551, L7
\bibitem[Berkhuijsen et al.(2006)]{ber06}
Berkhuijsen, E. M., Mitra, D., \& M\"{u}ller P., 2006, AN, 327, 82
\bibitem[Brentjens \& de Bruyn(2005)]{br05}
Brentjens, M. A., \& de Bruyn, A. G. 2005, \aap, 441, 1217
\bibitem[Brentjens(2011)]{br11}
Brentjens, M. A. 2011, \aap, 526, 9
\bibitem[Burkhart et al.(2012)]{bur12}
Burkhart, B., Lazarian, A., \& Gaensler, B. M. 2012, \apj, 749, 145
\bibitem[Burn(1966)]{bu66}
Burn, B. J. 1966, \mnras, 133, 67
\bibitem[Cordes \& Lazio(2002)]{cl02}
Cordes, J. M., \& Lazio, T. J. W. 2002, preprint (astro-ph/0207156)
\bibitem[de Bruyn \& Brentjens(2005)]{db05}
de Bruyn, A. G., \& Brentjens, M. A. 2005, A\&A, 441, 931
\bibitem[Farnsworth et al.(2011)]{far11}
Farnsworth, D., Rudnick, L., \& Brown, S. 2011, Astron. J, 141, 28
\bibitem[Frick et al.(2011)]{fri11}
Frick, P., Sokoloff, D., Stepanov, R., \& Beck, R. 2011, MNRAS, 414, 2540
\bibitem[Gaensler et al.(2005)]{gae05}
Gaensler, B. M., Haverkorn, M., Staveley-Smith, L., Dickey, J. M., McClure-Griffiths N. M., Dickel, J. R., \& Wolleben, M., 2005, Science, 307, 1610
\bibitem[Gaensler et al.(2008)]{gmcm08}
Gaensler, B. M., Madsen, G. J., Chatterjee, S., \& Mao, S. A. 2008, PASA, 25, 184
\bibitem[Gaensler et al.(2011)]{gae11}
Gaensler, B. M., Haverkorn, M., Burkhart, B., Newton-McGee, K. J., Ekers, R. D., Lazarian, A., McClure-Griffiths, N. M., Robishaw, T., Dickey, J. M., \& Green, A. J. 2011, Nature, 478, 214
\bibitem[Giacinti et al.(2010)]{gkss10}
Giacinti, G., Kachelrie{\ss}, M., Demikov, D. V., \& Sigl, G. 2010, Journal of Cosmology and Astroparticle Physics, 8, 36
\bibitem[Govoni et al.(2010)]{gov10}
Govoni, F., Dolag, K., Murgia, M., Feretti, L., Schindler, S., Giovannini, G., Boschin, W., Vacca, V., \& Bonafede, A. 2010, \aap, 522, 105
\bibitem[Heald et al.(2009)]{hea09}
Heald, G., Braun, R., \& Edmonds, R. 2009, \aap, 503, 409
\bibitem[Hill et al.(2008)]{hil08}
	Hill, A. S., Benjamin, R. A., Kowel, G., Reynolds, R. J., Haffner, L. M., \& Lazarian, A. 2008, \apj, 686, 363
\bibitem[Ideguchi et al.(2014)]{ide14}
Ideguchi, S., Takahashi, K., Akahori, T., Kumazaki, K., \& Ryu, D. 2014, PASJ, 66, 5
\bibitem[Kim et al.(1999)]{krjh99}
Kim, J., Ryu, D., Jones, T. W., \& Hong, S. S. 1999, \apj, 514, 506
\bibitem[Kumazaki et al.(2014)]{kum14}
Kumazaki, K., Akahori, T., Ideguchi, S., Kurayama, T., \& Takahashi, K. 2014, PASJ, in press (arXiv:1402.0612)
\bibitem[Machida et al.(2013)]{mac13}
Machida, M., Nakamura, K., Kudo, T., Akahori, T., Sofue, Y., \& Matsumoto, R., 2012, \apj, 764, 81
\bibitem[O'sullivan et al.(2012)]{osu12}
O'Sullivan, S. P., Brown, S., Robishaw, T., Schnitzeler, D. H. F. M., McClure-Griffiths, N. M., Feain, I. J., Taylor, A. R., Gaensler, B. M., Landecker, T. L., Harvey-Smith, L., \& Carretti, E. 2012, MNRAS, 421, 3300
\bibitem[Pizzo et al.(2011)]{piz11}
Pizzo, R. F., de Bruyn, A. G., Bernardi, G., \& Brentjens, M. A. 2011, \aap, 525, 104
\bibitem[Schnitzeler et al.(2009)]{sch09}
Schnitzeler, D. H. F. M., Katgert, P., \& de Bruyn, A. G. 2009, \aap, 494, 611
\bibitem[Sofue et al.(2010)]{smk10}
Sofue, Y., Machida, M., \& Kudoh, T. 2010, \pasj, 62, 1191
\bibitem[Sun et al.(2008)]{sun08}
	Sun, X. H., Reich, W., Waelkens, A., \& En{\ss}lin, T. A. 2008, \aap, 477, 573
\bibitem[Sun et al.(2014)]{sun14}
	Sun, X. H., Rudnick, L., Akahori, T. et al. (2014), submitted to AJ
\bibitem[Waelkens et al.(2009)]{wae09}
	Waelkens, A., Jaffe, T., Reinecke, M., Kitaura, F. S., \& En{\ss}lin, T. A., 2009, \aap, 495, 697
\bibitem[Wollenben et al.(2010)]{wol10}
Wolleben, M., Landecker, T. L., Hovey, G. J., Messing, R., Davison, O. S., House, N. L., Somaratne, K. H. M. S., \& Tashev, I. 2010, \aj, 139, 1681
\end{thebibliography}
\end{document}